# Accelerating MRI Colon Volume Measurements and Reducing Inter-Observer Variation through Automatic Segmentation and Human-in-the-Loop Correction

Abi Spicer*[1,2], Stephen Lloyd-Brown*[3], Neele Dellschaft[1,4], Luca Marciani[2,4], Robin Spiller[2,4], Maura Corsetti[2,4], Penny Gowland[1,2], Xin Chen[3], Caroline Hoad[1,4]

[1]Sir Peter Mansfield Imaging Centre, University of Nottingham, Nottingham, United Kingdom

[2]Nottingham Digestive Disease Centre, School of Medicine, University of Nottingham, Nottingham, United Kingdom

[3]Intelligent Modelling and Analysis, School of Computer Science, University of Nottingham, Nottingham, United Kingdom

[4] NIHR Nottingham Biomedical Research Centre BRC, Nottingham University Hospitals NHS Trust and the University of Nottingham, Nottingham, UK

*Joint first Author

## Plain Language summary (150 words)

The size of the colon relates to health, disease, and treatment. Currently, to measure colon size, a person must manually draw around (segment) the colon on an MRI scan. Current analysis methods introduce complications, such as the participant needing to take medicine that may interact with what we want to look at or add complex computing steps.

In our study, images of the colon were taken using an MRI scanner, with subjects having to take no extra medication. Machine Learning (ML) was used to segment the colon and volume measurements were obtained. Using ML trained on previous manual segmentations followed by manual corrections was over 80% quicker than manual segmentations and produced results which agreed with experts. ML agreement with experts was higher than the two experts' agreement with each other. This method is a big step towards being able to efficiently assess colon size and its relation to health.

## Abstract

## Background

The movement distribution, and volume of both chyme and gas in the colon, are important metrics to understand colonic function in health, disease, and the effects of treatments and different foodstuffs. Current methods available for assessment of these colonic contents using MRI consist mainly of manual segmentation or semi-automatic segmentation. However, these methods of segmentation are very labour intensive and too slow for clinical applications, require expert knowledge and some semi-automatic methods require use of bowel preparation.

## Methods

MRI scans were acquired in 2 breath holds using mDIXON sequences. We used the “No New U-Net” (nnU-Net) ML model to automatically segment the colon, including colonic regions (ascending, transverse, descending and sigmoid-rectal). The ML-generated masks were corrected manually and the time taken for correction was recorded. ML segmentations were compared to both manual segmentations and observer corrected ML (CorrML) segmentations. Observer repeatability was also evaluated for both manual and CorrML methods to create a benchmark for the allowable error in the automatic segmentations.

## Results

Analysis time was significantly reduced ($p<0.0001$) from 56 mins (±11 mins (SD)) for manual masks to 11 mins (±5 mins (SD)) for CorrML masks. Both DICE and ICC values showed excellent agreement between manual, ML and CorrML segmentations for whole colonic volume (ICC = 0.96) whilst regional volumes were good-excellent (ICC = 0.80-0.95). Inter-observer repeatability was improved when using CorrML methods over manual segmentation (ICC manual > 0.89, CorrML > 0.93).

## Conclusions

Analysis time was reduced by over 80% when using CorrML methods and whole colonic volumes measured by ML would be suitable for use with minimal checks. Hence the methods proposed here would be clinically useful.

**Highlights ( 3 to 5 bullet points)**

- A new fast method for segmenting an unprepared colon from MRI

- Whole colonic volume measured by ML showed less variation with matched manual segmentations than the variations seen between two observers manually segmenting the same colon
- This could move colonic volume assessment from a research tool to a clinical assessment

# 1 Introduction

The movement of chyme, its distribution and volume in the colon as well as the volume of gas are all important metrics to understand colonic function in health, disease, and the effects of treatment. The assessment of colonic content has previously been conducted using computer tomography (CT) though this subjects participants to large doses of ionising radiation (1–3). Magnetic Resonance Imaging (MRI) has no ionising radiation, and is therefore a safer alternative for non-invasive assessment of the colon in both healthy volunteers and patients.

Measuring the colonic volume using MRI scans has a wide range of applications for both health and disease (4). Previous literature has shown colonic volume and gas assessment are valuable in giving insight into Irritable bowel syndrome (IBS). Thus while the physiological responses to FODmaps (Fermentable Oligosaccharides Disaccharides Monosaccharides And Polyols) such as inulin are similar for both IBS patients and healthy participants in patients the peak symptom intensity of bloating after inulin correlated with colonic gas (5). Recent studies of constipated patients showed that those with a colonic volume > the 90$^{th}$ centile of normal (923 ml) had a reduced postprandial manometric response, reduced response to macrogol challenge and reduced pain on bisacodyl treatment(6). Other studies have also explored colon volumes in studies of lactose malabsorption (7–9), cystic fibrosis (8,9), and effects of both food (10–12) and drugs in the lower GI tract (13,14). However, colonic volume and colonic gas assessment is underutilised and difficult to translate clinically due to the time-consuming nature of manual (or semi-automated) segmentation currently required (15,16).

Segmentation, the process of drawing around an organ of interest to create a mask, is complex for the colon. There are large changes in colon geometry within and between individuals and on MR images there is variable contrast both within the lumen and with surrounding tissue as well as gas accumulation in the lumen. The gold standard for colon volume assessment has been manual segmentation, but this is not entirely objective, requires a trained observer, and is time consuming. The properties which make colon segmentation difficult for trained observers also make the task challenging for classical automatic segmentation methods. Variable contrast makes thresholding inappropriate, and the usefulness of more sophisticated methods, such as active contours is limited in the case of collapsed regions, such as the descending colon, or regions with poorly defined boundaries such as the sigmoid colon, requiring frequent intervention from the user throughout the segmentation process (17).

Recent advances in ML have allowed for automatic segmentation of many organs in the body (18–21). One solution publicly available is TotalSegmentator (22), however, given the wide scope of TotalSegmentator, the authors did not report an exhaustive evaluation of the segmentation success for all organs, making it hard to predict its success on local datasets.

Semi-automatic segmentation of the colon has been investigated, though this still required users to draw ROIs on all images where the colon was present (16). Fully automatic methods of segmenting the GI tract, including the colon, have generally not evaluated individual organs (23–27), have required bowel preparation (1600–2000 mL of 2.5% mannitol solution) (28) or required extensive knowledge of machine learning to implement (29). Ideally, a segmentation method would allow the user to submit all scans to an easily trainable, publicly available segmentation pipeline, and subsequently check and correct masks if necessary.

In line with other MR segmentation studies (30–32), we utilised the popular network "No New U-Net" (nnU-Net) as our ML segmentation model. The network uses convolutional filters to highlight regions of interest. Unlike traditional image processing techniques, these filters are "learnt" by the network, rather than being manually determined, by providing example segmentations for the network to learn from. For large images, particularly in 3D, images are often be split into patches, so that the network is kept small enough to run on available hardware. This means the network only learns from smaller localised regions, rather than a "big picture" overview. Consequently, these networks are good at segmenting local differences in contrast, but can struggle in applications where region boundaries are less clearly defined using contrast, such as anatomical brain segmentation. We use nnU-Net for three reasons. First, because U-Net (33) models demonstrate excellent performance in medical image segmentation (34–36). Second, because nnU-Net can optimise its own hyperparameters, removing the capacity for human error in tuning (a key concern in ML-based research) as a factor in performance. Third, it is publicly available to use.

The goal of this study was to create an nnU-Net model capable of segmenting the colon automatically and evaluate the model through comparisons with ground truth masks and manually corrected ML-generated masks. Results were compared to inter-observer repeatability for both ground truth and corrected ML (CorrML) masks to define maximal accepted difference between the CorrML and ground truth annotations.

# 2 Methods

## 2.1 Data Acquisition

As part of a larger study, 22 participants visited the Sir Peter Mansfield Imaging Centre three times for a three-way randomised intervention study (37). This study was carried out with institutional ethical approval (University of Nottingham FMHS 328-0723, approval given on 24$^{th}$ August 2023) and was preregistered before commencement (clinicaltrials.gov/study/NCT06291272). All participants gave consent for their data to be used in retrospective studies. 19 retrospective datasets were retrieved for use in this study, demographics were 10 male and 12 female, aged 28 ± 9 yrs and BMI = 24.6 ± 4 kgm$^{-2}$.

At each visit, participants arrived fasted and were then fed inulin mixed with maltodextrin (drink), methylcellulose, or psyllium (both as gels), which created different contrasts in colonic chyme (Figure 1) and varying levels of gas, adding a level of complexity to the dataset. MRI scans were performed at baseline (BL), T = 0 (immediately after consuming the meal), and then hourly over a 6-hour period and finally a single scan at T = 24 hours on a 3T Philips wide bore Ingenia scanner (Best, The Netherlands). At each time point a dual echo mDIXON (field of view = 250x375, 2 stacks, 30mm overlap, echo time = 1.12/3.60ms, repetition time 3.2ms, flip angle 10°, reconstructed voxel size 0.98x0.98x1.80 $mm^3$) images were obtained for colonic volume measurements (38). Four different contrasts were generated from the mDIXON acquisition: water-only, fat-only, in-phase, and out-of-phase. Each contrast of the mDIXON scans was re-sampled to a matrix of 0.98x0.98x5.4 $mm^3$, to reduce the number of slices needed for analysis, ending up with 37 slices of 5.4 mm thickness, with 15-25 containing the colon.

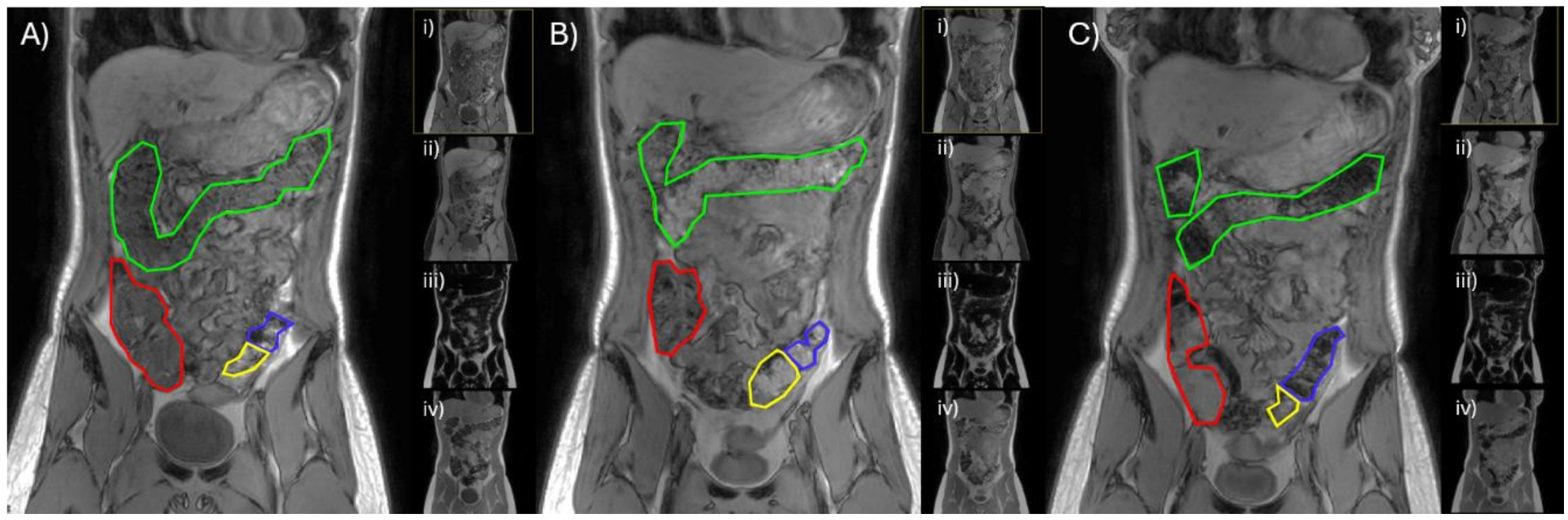

*Figure 1 - Images from the same time after intervention for the same participant across the 3 visits (A, B, C). Images i – iv are the different contrasts produced by the Dixon scans. Ground truth masks used to show colon regions. Red – Ascending, Green – Transverse, Blue – Descending and Yellow – recto-sigmoid colon.*

## 2.2 Segmentation Mask Generation

The colon in the images was then segmented by three different processes: manually by four trained observers (ground truth), by Machine Learning (ML) model, and by ML model with subsequent manual correction by two of the observers (CorrML). For both ground truth and CorrML methods the colon was segmented into 4 regions: Ascending (AC), transverse (TC), descending (DC) and rectal-sigmoid (SC). Each observer was given the following guidance for region definitions: the AC begins at the cecum (excluding the terminal ilium), and ends at the highest point of the hepatic flexure (where the colon begins to loop over itself), from this point the TC begins, and ends at the highest point of the splenic flexure, here the DC begins, and ends at the pelvic brim where the colon begins to change direction, and extends distally, this is the start of the sigmoid colon which extends to the rectum which in turn terminates at the anus. The sigmoid and rectum were combined into one region (SC).

### 2.2.1 Manual masks

As part of preliminary analyses in the primary study to assess the effects of the intervention, 4 observers manually segmented data from T=baseline (BL), 180, 240, 300, and 360 minutes after the intervention. Manual segmentations were all done using MIPAV software (39), by manually tracing a polygon region of interest segmentation tool on the 2D images of the colon (15), with each region of the colon having its own label. This gave a mask that had 5 labels, 0 = background (in this case anything that is not the large bowel), 1 = AC, 2 = TC, 3 = DC, 4 =SC.

Upon completion of data collection in the primary study the remainder of the timepoints (T = 0, 60, and 120 minutes, and 24h post intervention) required analysis. For 9 randomly selected participants, these remaining timepoints were analysed using the same manual analysis methods, which combined with the masks from T = BL, 180, 240,300 and 360 for these 9 participants created 243 pre-annotated masks, which were used as the training data in our AI model. The AI model was then used to process the data of all timepoints from the remaining 10 participants. The initial manual analysis of timepoints T = BL, 180, 240, 300, and 360 minutes these 10 participants were kept as a separate test set of 150 scans, to be used in evaluating the AI models, Figure 2.

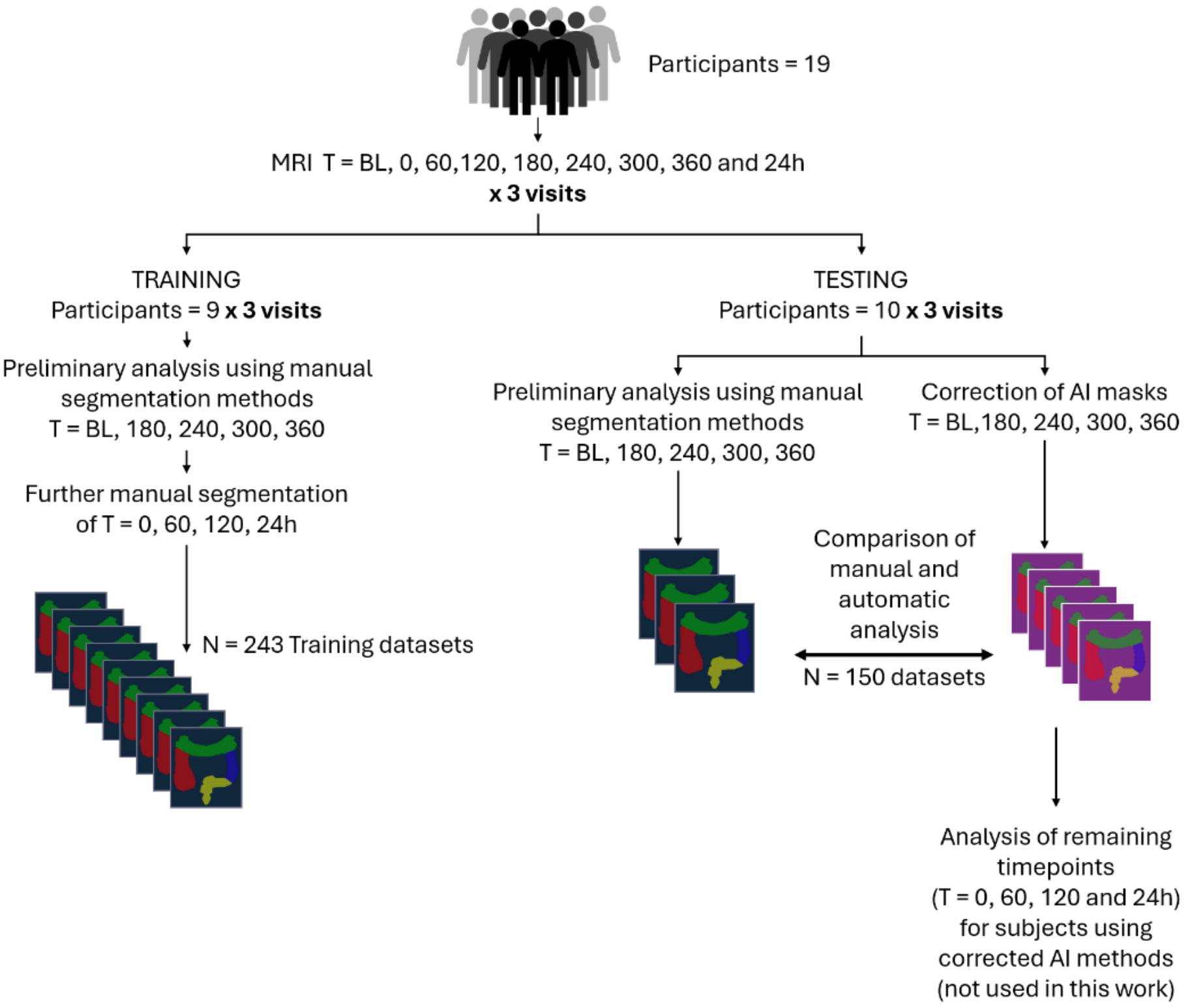

*Figure 2 - Distribution of data into training and testing data for AI models*

## 2.2.2 ML model masks

The nnU-Net model was trained using all 4 different contrasts produced by the mDIXON data. During training, the model simultaneously learns from the "Whole Colon" label, as well as learning from the labels of the colonic regions (AC, TC, DC, SC). (Figure 1). After training, the model produces the "Whole Colon" label first, which is then immediately overwritten by the more specific labels of the constituent regions (AC, TC, DC, SC) prior to output. The result is five labels, consisting of the four regions, as well as any identified colon which fails to be attributed to a specific region. After masks for all testing data were produced, timestamps were viewed in order to determine the time taken for masks to generate.

## 2.2.3 Corrected ML mask

The masks generated by the nnU-net model were edited by two of the four observers who had created the respective ground truth masks for these subjects (>9 months earlier). The selected observers had different levels of experience with colon segmentation and different specialities (observer A has an MR physics background and was new to the analysis, while observer B has a physiology background and 9 years' experience in colonic segmentation). Prior to presenting the ML masks to observers, the multiple regions produced by the model were combined into a single label.

This is because it was assumed that asking the observer to split the region would help maintain engagement, see section 4. Each observer opened the mask in ITK-SNAP (40) overlayed on the mDIXON data. The scalpel or brush tool were used to divide the mask into 4 regions, and to correct any areas of missing or erroneously labelled colon, Figure 3.

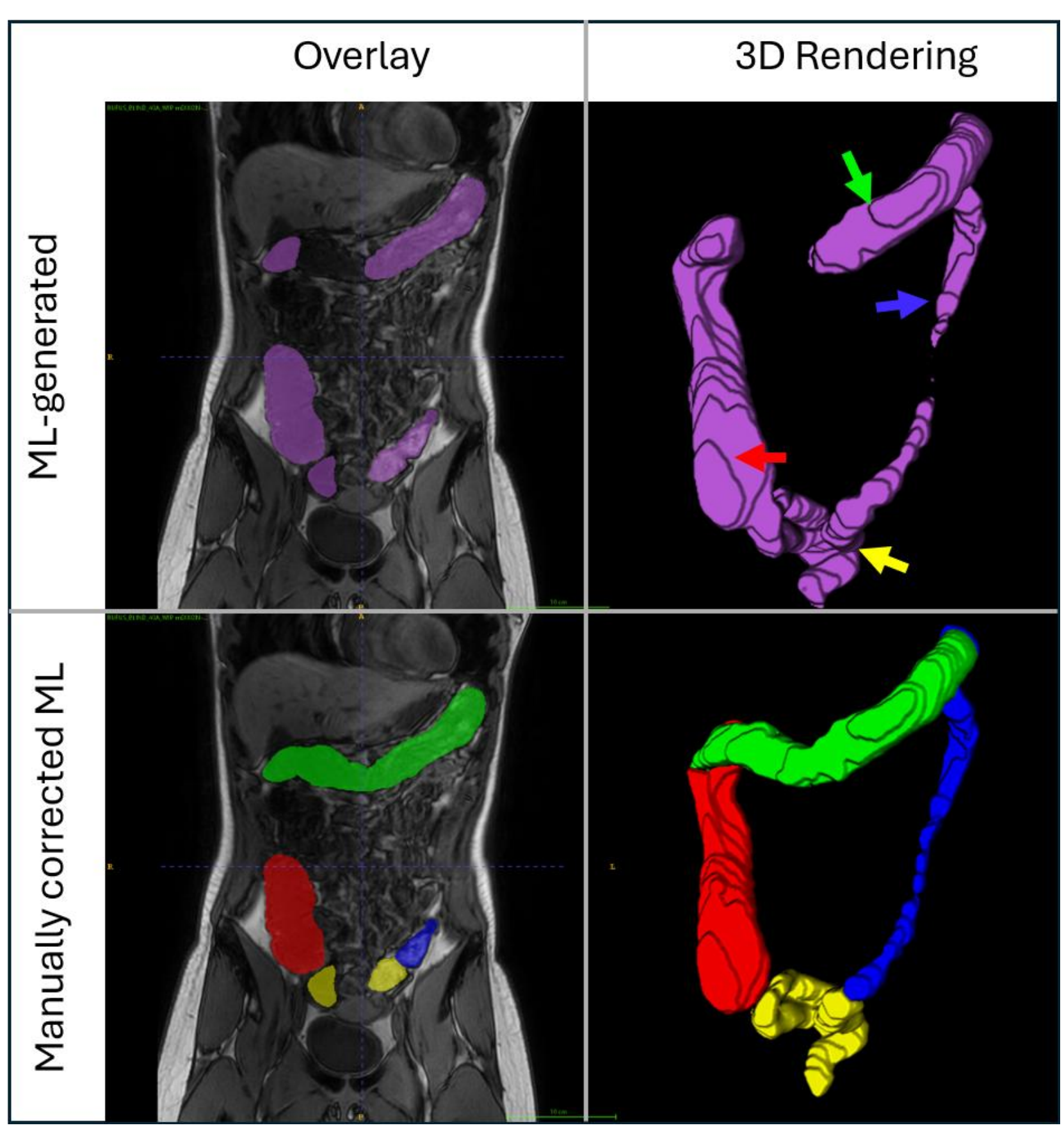

*Figure 3 - Example of a machine learning (ML) generated mask with the respective corrected ML mask. Arrows and colon segments labelled are colour coded Red – ascending, Green – transverse, Blue – descending and yellow – recto-sigmoid colon.*

## 2.3 Assessing the accuracy of the Masks

### 2.3.1 Variability between manual, automatic and corrected masks

Whole colon and region masks for the ground truth, Machine Learning (ML) generated and corrected ML (CorrML) mask were compared using Dice scores, a widely used segmentation metric (31,34–36).

Volume differences were compared using scatter plots with linear regression (GraphPad Prism Version 10.6.0, GraphPad Software, San Diego, California, USA) and a two-way mixed intra-class correlation coefficient (ICC) with absolute agreement was used to compare (IBM SPSS, Version 29, IBM Corp., 2023). Difference masks of the whole colon were generated to visualize where changes occur between ground truth and ML or CorrML masks.

### 2.3.2 Inter-observer repeatability

For inter-observer repeatability, there was a pool of 120 datasets available, consisting of scans from 8 participants. It was decided that 30 segmentations would strike the appropriate balance between sufficient data to evaluate inter-observer variations, and staff time to create the masks. Initially 30 unique datasets were selected for manual segmentation by both observers (AS and ND). Following this, from the remaining pool of data, a separate set of 30 unique scans had their ML masks corrected (CorrML) by both observers. The observers recorded how long each manual segmentation and correction of each ML mask took.

Dice scores were used to compare differences in the masks between observers A and B, and a two-way mixed ICC with absolute agreement was used to compare differences in volumes for both manual and CorrML analysis between observers. Volumes from each method were compared using linear regression and Bland-Altman's plots.

# 3 Results

## 3.1 Creating the Colon Segmentation Masks

Per 37 slice scan covering the abdomen, ground truth masks took 56 mins (±11 mins (SD)) to create, whilst ML masks took 6s to generate. Correcting ML masks took an average of 11 mins (±5 mins (SD)) per scan which was significantly faster than creating ground truth masks ($p < 0.0001$, Welch's unpaired t-test, GraphPad Prism Version 10.6.0, GraphPad Software, San Diego, California, USA).

## 3.2 Variability between manual, automatic and corrected masks

Mean Dice scores comparing ground truth, ML and CorrML whole colon masks were all >0.89, and for colonic regions, mean Dice scores were all >0.82 (Table 1).

*Table 1 - Dice scores for comparisons between the Ground Truth (GT), AI generated (AI) and Corrected AI (CorrAI) segmentation results (mean ± SD)*

| | Region | Total | | AC | | TC | | DC | | SC | |
|---|---|---|---|---|---|---|---|---|---|---|---|
| Dice Scores | GT – AI | 0.89 | ±0.03 | 0.86 | ±0.09 | 0.83 | ±0.08 | 0.82 | ±0.09 | 0.83 | ±0.07 |
| | AI – CorrAI | 0.98 | ±0.02 | 0.94 | ±0.09 | 0.92 | ±0.06 | 0.91 | ±0.08 | 0.94 | ±0.06 |
| | GT – CorrAI | 0.90 | ±0.03 | 0.89 | ±0.04 | 0.86 | ±0.06 | 0.86 | ±0.05 | 0.86 | ±0.06 |

Inter-method volume agreement measured using ICC for total colon volume was excellent (> 0.95). Comparison of regional volumes between ground truth and ML were good-excellent (ICC ranged from 0.80-0.92), whilst ground truth to CorrML was excellent (all ICC values >0.92). Comparisons between ML and CorrML were also good-excellent (ICC ranged from 0.85 - 0.95), Table 2.

*Table 2 - ICC values for volume comparisons between different methods of segmentation*

| | Regional | Total | AC | TC | DC | SC |
|---|---|---|---|---|---|---|
| ICC | GT – AI | 0.95 | 0.85 | 0.84 | 0.80 | 0.92 |
| | GT – CorrAI | 0.95 | 0.94 | 0.92 | 0.93 | 0.94 |
| | AI – CorrAI | 0.99 | 0.91 | 0.89 | 0.85 | 0.95 |

Volumes of the whole colon ground truth masks tended to be slightly lower than the whole colon volumes measured by either ML or CorrML methods. R-values from linear regression for total colon volume of ground truth versus ML and CorrML masks were all >0.95 (Figure 4). R-values from comparing regional volumes between ground truth and ML masks were low, with values ranging from 0.07 to 0.69, Figure 4. Whilst R-values comparing region volumes from ground truth to CorrML were all >0.94 (Figure 4). Figure 5 shows example difference masks of ground truth to ML and CorrML masks, showing typical errors. Regions where the ML masks failed, such as a collapsed DC and area of the TC can be seen to be fixed in the observer-corrected mask. The area within the AC that is present in the ground truth but not in the ML or corrML masks, correlates to the terminal ilium when overlayed on the MR image.

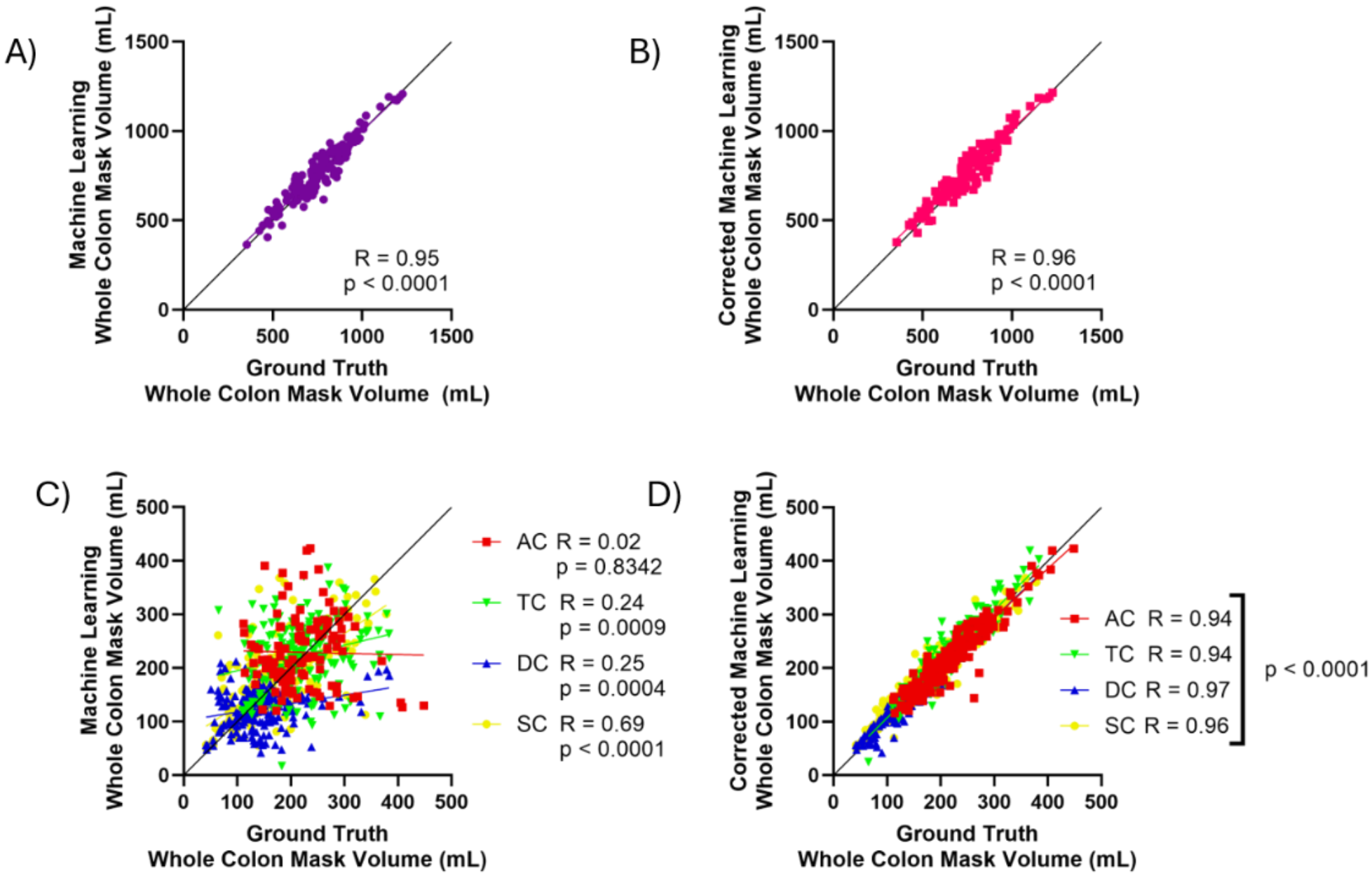

*Figure 4 - Whole colon volumes measured by ground truth masks plotted against (A) AI and (B) corrected AI masks. C) AI and (D) Corrected AI volumes plotted against ground truth for the ascending (AC), transverse (TC), descending (DC) and recto-sigmoid (SC) colon regions separately. All graphs combine results from both observers*

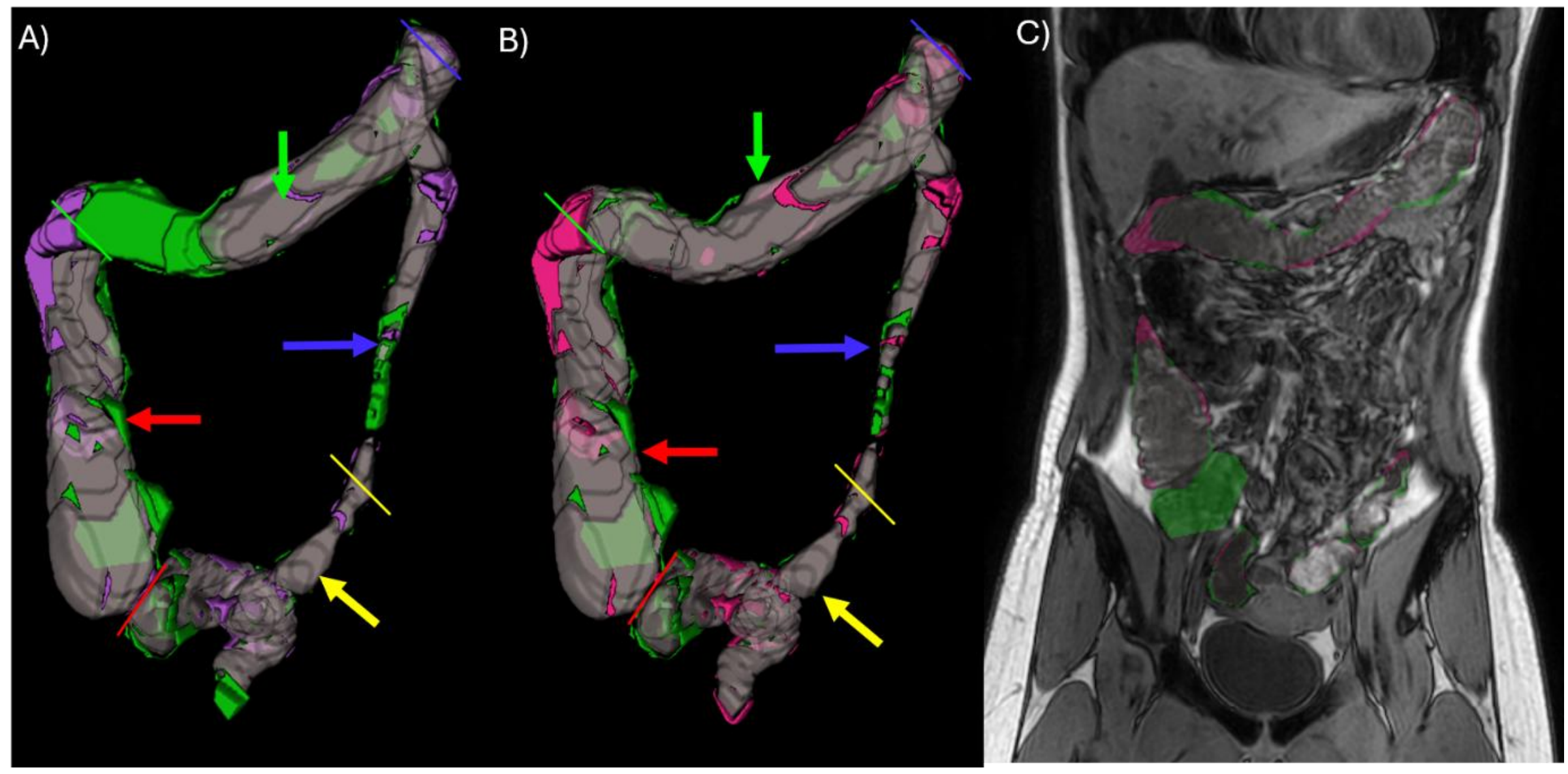

*Figure 5 - Example difference masks comparing manual ground truth masks with A) Machine Learning (ML) generated and B) corrected ML (same scan). The green areas are only present in the ground truth mask but not the mask it is being compared to, and purple/pink areas are only present in the ML/ Corrected ML mask respectively. Colonic regions ascending, transverse, descending, and recto-sigmoid colon are identified with red, green, blue and yellow arrows respectively with same-coloured lines showing approximate transition points from one region to the other. C) shows the difference mask shown in B) overlayed on the mDIXON image.*

## 3.2.1 Inter-observer repeatability

*Table 3 - Dice scores for comparisons between Rater A and Rater B for the (n =30) Ground Truth (GT) and (n = 30) Corrected Machine Learning (CorrML) masks*

| | Colon Region | Total | | AC | | TC | | DC | | SC | |
|---|---|---|---|---|---|---|---|---|---|---|---|
| REP Dice | (n = 30) GT | 0.88 | ±0.03 | 0.88 | ±0.04 | 0.86 | ±0.05 | 0.86 | ±0.05 | 0.83 | ±0.07 |
| | (n = 30) CorrAI | 0.98 | ±0.02 | 0.96 | ±0.06 | 0.95 | ±0.04 | 0.93 | ±0.07 | 0.96 | ±0.05 |

*Table 4 - ICC values for volume comparisons between Rater A and Rater B for total colon volume and regional mask measured using Ground Truth (GT) and Corrected Machine Learning (CorrML) methods (n=30 both groups)*

| | Colon Region | Total | AC | TC | DC | SC |
|---|---|---|---|---|---|---|
| REP ICC | GT | 0.93 | 0.94 | 0.95 | 0.89 | 0.91 |
| | CorrAI | 0.99 | 0.97 | 0.93 | 0.95 | 0.98 |

Mean Dice scores were >0.83 and >0.93 for manual and CorrML analysis respectively (Table 3), with ICC values of >0.89 and >0.93 for manual and CorrML respectively (Table 4). Figure 6 shows the volume comparisons for observer A and B for both ground truth and CorrML with R values >0.95 for total colon volume and all regional volumes >0.93. The Bland-Altman plot shows that more variation between observers is seen in ground truth masks compared to the CorrML methods with the DC and SC generally being most variable between observers. Example difference masks for 2 participants are shown in Figure 7; one had high variation, and the other low variation between observers for both ground truth and CorrML methods. Fewer differences are seen in the CorrML compared to the ground truth masks.

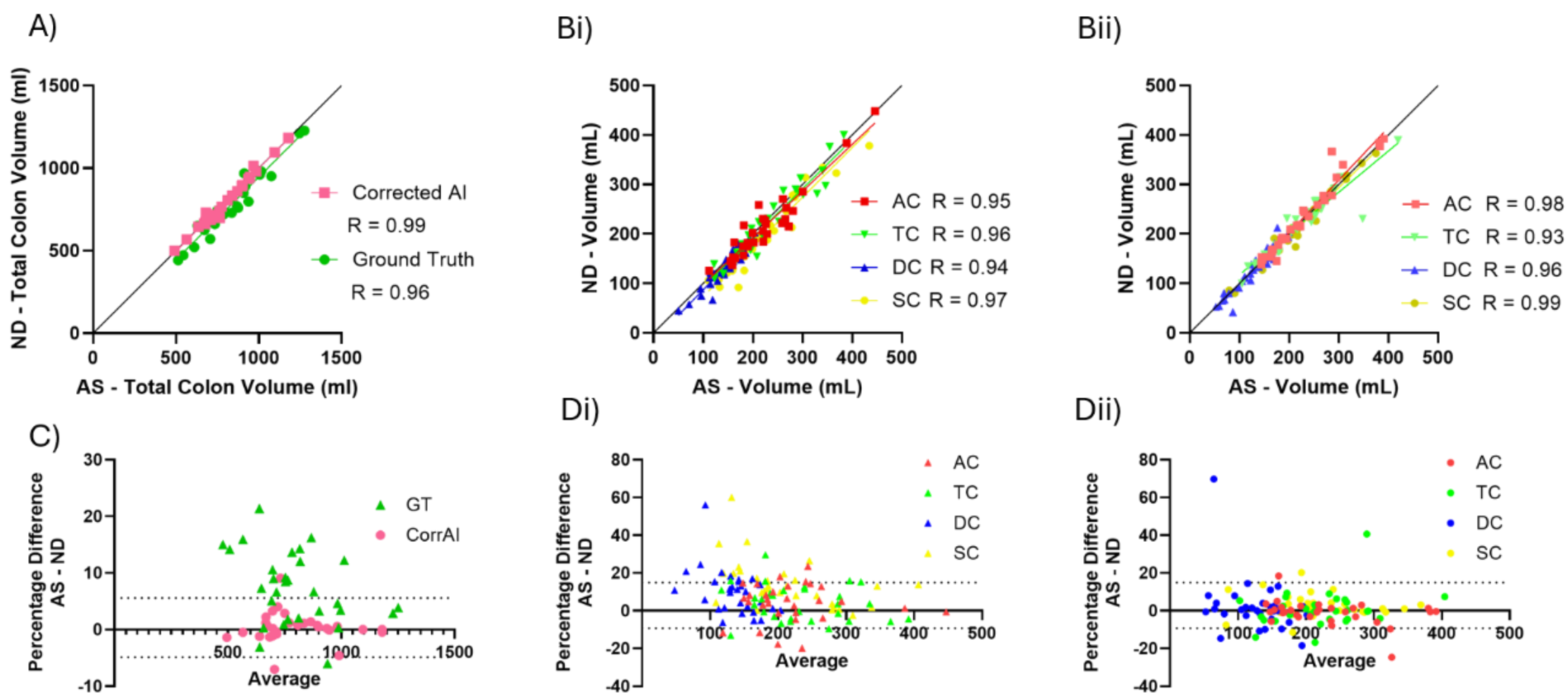

*Figure 6 - Comparisons of volumes measured by both observers for Ai) total colon and Ground Truth (GT) and Corrected Machine Learning (CorrML) and B) regional volumes for i) GT and ii) CorrML masks all with linear regression along with respective Bland-Altman plots showing the percentage difference ((AS-ND)/Average) compared to the average volume measured by the observer for C) the whole colon and D) regional volumes.*

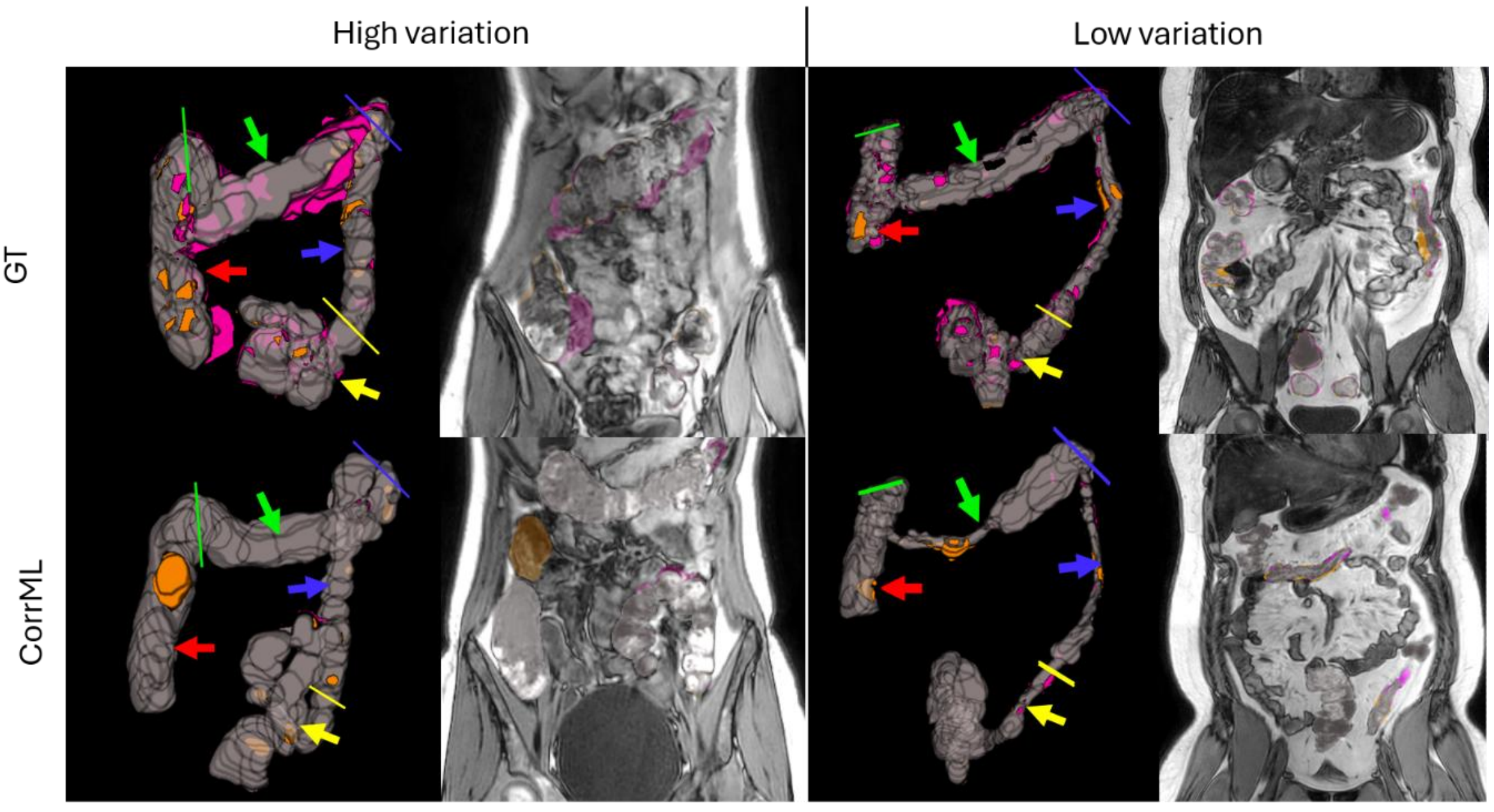

*Figure 7 - Example difference masks comparing between observers for a Ground Truth (GT) (top) dataset and a different Corrected Machine Learning (CorrML) (bottom) dataset. Pink shows areas only present in AS's mask and orange shows areas only present in ND's mask. Each column is one participant but due to dataset selections timepoints are different for the GT versus the CorrML comparisons. The left masks show a relatively high variation between observers, with the middle showing these difference masks overlayed on the respective image. The right shows example masks with relatively low variation between observers. Colonic regions AC, TC, DC, and SC are identified with red, green, blue and yellow arrows respectively with same-coloured lines showing approximate transitions from one region to the other.*

# 4 Discussion

## 4.1 Creating the Colon Segmentation Masks

As stated in section 2.1, there were concerns about maintaining observer engagement when correcting erroneously labelled colon in the ML masks. It was believed if a mask appeared correct at first glance, users would not take the extra time to carefully inspect all slices. Hence, the multiple region masks produced by the ML model were combined to represent the total colon.

As expected, the process of correcting an ML mask was significantly faster than the manual segmentation for ground truth masks.

## 4.2 Assessing the Masks

Overall, the ML successfully segmented the colon, despite missing some areas (Figure 3) or labelling other parts of the body as colon. Dice scores presented in this work for ground truth to ML are in line with those previously reported for whole GI tract segmentations and the colon using UNet, a related method where the user has to set the optimisation parameters (Dice 0.71 – 0.91 (41–43)) and are in line with nn-UNet segmentations of the pancreas (Dice > 0.75, (44)).

A larger similarity was seen in the Dice scores comparing ML and CorrML masks, compared to the smaller Dice score for ground truth to CorrML/ML masks. The subtraction image shown in Figure 5 shows that the difference in the ground truth to ML/CorrML masks often occurs at the edges, which is likely due to the different method of creation. ground truth masks were created in MIPAV using the polygon tool which, due to the nature of a finite polygon, can lead to jagged edges. Conversely, the ML often has continuous, smooth edges which are carried into the CorrML masks. These smooth edges masks are likely the source of the slightly higher volumes measured from the ML and CorrML compared to the ground truth.

Comparison of whole colon volumes made using ICC scores showed excellent agreement across all methods. For regional mask volumes the TC and DC consistently had more variation between methods, likely due to both these regions having two flexures. Flexure points of the colon are difficult to define and are a very variable part of colonic volume analysis, hence having no single well defined boundary will allow for more variation in volume to occur.

### 4.2.1 Inter-observer

There was little variation in the masks or volumes measured between two observers, and in general, across all metrics, less variation was seen when CorrML methods were used. Dice similarity scores showed high inter-observer agreement between for both ground truth and CorrML masks and were similar to those reported by Sandberg et al., 2015 for their semi-automatic segmentation methods (16). Mean inter-observer Dice scores for CorrML exceeded that of inter-observer ground truth Dice scores for total and regional masks (Table 3). ICC scores show excellent agreement between observers for total colonic volume and all regional volumes for both ground truth and ML corrected methods.

The TC is the only region where the volumes in the ground truth masks had higher agreement than in the CorrML, for both ICC (0.95 and 0.93, respectively) and R values (0.96 and 0.93, respectively). This may be due to difference in methods between ground truth mask creation (drawing polygons on 2D images) and ML mask editing (3D render segmentation). The colon often folds over itself, especially at flexures. The TC starts at the hepatic, and ends at the splenic flexure, meaning unlike other colonic regions, it has 2 points which are likely to be folded over itself. The 3D visualisation of the colon in ITKSnap allows observers to easily see how the colon folds over itself, and can be segmented using the scalpel tool in 3D. Conversely, in MIPAV, the observer is often left to pick an arbitrary slice where the colon appears to be at the highest point in 2D to mark this as a transition, even if, due to the way the colon folds, the area may be better defined as 2 colonic regions. In 2D there are fewer degrees of freedom to introduce variation between observers in the TC flexures, whilst 3D segmentation allows for better visualisation and more degrees of freedom, increasing variability.

The difference between the observers, often occurred at the edges of the colon. When the difference masks are overlayed on the respective images it can be seen that the regions in observer AS's mask which are not in observer ND's mask are areas of partial voluming, which is where the thickness of the slice is bigger than amount of colon in that slice and this is a subjective area of inclusion in colonic volumes. The consistently higher volumes measured by AS are also likely a result of consistent inclusion of partial voluming regions. The difference between observers is reduced to around 8% for CorrML masks and there is less variation in the difference between the observers. This may be due the ML doing a 'good' to 'excellent' job (as determined by Dice and ICC scores) of identifying the colon initially. As discussed earlier, the edges of the colon are often well defined in the ML masks, needing very little manual correction and hence very little sources of variation between observers.

The larger variations in regional volumes for ground truth methods compared to CorrML is likely a result of the differences in segmentation methods discussed above. Observers tended to agree on colonic region transition points as shown by the high correlation in volumes between observers (R > 0.93 for all regions). However, we observe the percentage difference between observers was higher than that for total colon volume (up to 60% instead of 20% in ground truth), and that this difference was generally reduced with the use of CorrML methods, Figure 6. The subjectivity of the transition points has not changed between methods, however the method of segmentation has, showing that the use of a 3D segmentation software likely contributed to the reduced variation.

### 4.2.2 Summary

ML without correction appears to be suitable to measure whole colon volume but not regional volumes. A higher ICC value was seen when comparing whole colon volumes from the three analysis methods (ground truth to CorrML), compared to the ICC comparing two observers ground truth whole volumes (0.95 and 0.93, respectively). This trend is also seen in the plots for whole colon volume. R values for the ML masks without correction are similar to that between observers for the ground truth methods (R = 0.95 and R = 0.96, respectively) showing less or similar variability between methods than between observers depending on the metric used. Regional volumes ICC values were on average 0.11 points lower than the respective ICC value for comparisons between observers (both ground truth and CorrML). Similarly, R values of comparisons between ground truth with ML mask volumes were consistently much lower than the respective region R values seen from comparison between observers.

## 4.3 Strengths and limitations

MRI image data required for this method can be acquired in two breath-holds, making it easy to add to existing GI MRI protocols and therefore ideal for clinical applications. The model requires minimal preprocessing of the images, only requiring assembling of the 2 stacks (upper and lower abdomen).

Good versatility of the model was seen, with the model having been used on data with varying chyme contrast. Since the completion of this project, ongoing studies with different interventions are successfully implementing the model without further training, including data acquired on a 3T GE scanner, with different image resolution and slice thickness.

Training data was from a single study with only nine individuals, however this was mitigated due to contrast changes, and substantial colon volume changes from the interventions (27 scans per participant). The ability for the model to segment colons with large amounts of gas (>100ml) cannot be determined as there was only 1 dataset fitting this criterion in the testing data (37). However, this dataset had Dice scores for ground truth to ML comparisons of 0.88 for total colon volume and regional Dice scores all > 0.84 showing good agreement between the ML and ground truth masks.

Errorless automatic analysis was not achieved in this study. Mistakes in the ML masks are readily found upon visual inspections, and all masks in this dataset required some level of manual correction. This is more important for applications which may use the regional volume information for analysis of mass movements of chyme (45). Colonic masks created for volume analysis are often applied to other data for colonic gas measurements. The ML masks prior to correction would likely not be suitable for computing gas volumes within the colon, as colonic gas volumes tend to be only a small proportion of the total volume, and any erroneously labelled colon could miss larger proportions of the gas, which often lies at the anterior edges of the AC, TC and DC due to gravity when lying in the supine position (5).

Correction of the ML generated NIfTI masks was not possible in the software used for generating the polygon ground truth masks, hence this was carried out in ITKSnap. Rather than also moving to making the ground truth masks in the new software for this study, it was felt an authentic comparison would be achieved by using the established ground truth mask generation pipeline. The use of 2 different pieces of software will inevitably have added to any variations seen between the ground truth and CorrML masks, however it is very likely these variations are small in comparison to those introduced by having an existing mask presented to an observer compared to a blank image.

The model struggled with identifying the boundaries between the colonic regions. This is due to several factors but can be primarily attributed to the inherent limitations of nnU-Net, and convolutional neural networks (CNNs) more generally. The network is purely visual and relies on differences in local contrast to make voxel classification decisions. As the colon is contiguous and any changes to chyme texture are gradual, the network would be required to find its context solely from surrounding organs and tissue, which aren't consistent markers for colonic region boundaries. We believe the issue discussed would not be addressed by simple utilisation of additional data and an alternate method or continued manual input is required.

# 5 Conclusions

Overall, this study evaluated the suitability of the ML for segmentation of the colon segmentation of the colon for use in clinical practice. Analysis time was reduced by over 80% for the corrected ML methods compared to full manual annotation. Volume measured for whole colon from the ML prior to correction had excellent agreement with ground truth masks and would be suitable to present with minimal checks making it an excellent tool to help transition colonic volume assessment to the clinic. The model was unable to adequately determine colonic region boundaries, hence when regional masks are required observers will need to edit the mask provided to them. Future work should explore additions to this pipeline to allow for automatic regional segmentations.

# 6 Acknowledgments

The authors would like to acknowledge Alaa Alhasani, Joshua Reid and all the collaborators on the ENERGISE grant for collecting the data of the study the colon segmentations were taken from.

# 7 References

1. Barba E, Sánchez B, Burri E, Accarino A, Monclus E, Navazo I, et al. Abdominal distension after eating lettuce: The role of intestinal gas evaluated in vitro and by abdominal CT imaging. Neurogastroenterology and motility. 2019 Dec 1;31(12). doi:10.1111/NMO.13703 PubMed PMID: 31402544.

2. Barba E, Quiroga S, Accarino A, Lahoya EM, Malagelada C, Burri E, et al. Mechanisms of abdominal distension in severe intestinal dysmotility: abdomino-thoracic response to gut retention. Neurogastroenterology and motility. 2013 Jun;25(6). doi:10.1111/NMO.12128 PubMed PMID: 23607758.

3. Accarino A, Perez F, Azpiroz F, Quiroga S, Malagelada JR. Intestinal gas and bloating: effect of prokinetic stimulation. Am J Gastroenterol. 2008 Aug;103(8):2036–42. doi:10.1111/J.1572-0241.2008.01866.X PubMed PMID: 18802999.

4. Major G, Pritchard S, Murray K, Alappadan JP, Hoad CL, Marciani L, et al. Colon Hypersensitivity to Distension, Rather Than Excessive Gas Production, Produces Carbohydrate-Related Symptoms in Individuals With Irritable Bowel Syndrome. Gastroenterology. 2017 Jan 1;152(1):124-133.e2. doi:10.1053/j.gastro.2016.09.062 PubMed PMID: 27746233.

5. Major G, Pritchard S, Murray K, Alappadan JP, Hoad CL, Marciani L, et al. Colon Hypersensitivity to Distension, Rather Than Excessive Gas Production, Produces Carbohydrate-Related Symptoms in Individuals With Irritable Bowel Syndrome. Gastroenterology. 2017 Jan 1;152(1):124-133.e2. doi:10.1053/j.gastro.2016.09.062 PubMed PMID: 27746233.

6. Wilkinson-Smith V, Scott M, Menys A, Wiklendt L, Marciani L, Atkinson D, et al. Combined magnetic resonance imaging, high resolution manometry and a randomised trial of bisacodyl versus hyoscine shows the significance of an enlarged colon in constipation: the RECLAIM study. Gut (2024) (In press) [Internet]. 2024 Sep 23 [cited 2026 Feb 9]. Available from: https://gut.bmj.com/

7. Gunn D, Topan R, Fried R, Holloway I, Brindle R, Hartley S, et al. Efficacy and Mechanism Evaluation Ondansetron for irritable bowel syndrome with diarrhoea: randomised controlled trial. 2023. doi:10.3310/YTFW7874

8. Yule A, Sills D, Smith S, Spiller R, Smyth AR. Thinking outside the box: a review of gastrointestinal symptoms and complications in cystic fibrosis. Expert Rev Respir Med. 2023 Jul 3;17(7):547–61. doi:10.1080/17476348.2023.2228194 PubMed PMID: 37345513.

9. Ng C, Dellschaft NS, Hoad CL, Marciani L, Ban L, Prayle AP, et al. Postprandial changes in gastrointestinal function and transit in cystic fibrosis assessed by Magnetic Resonance Imaging. Journal of Cystic Fibrosis. 2021 Jul 1;20(4):591–7. doi:10.1016/J.JCF.2020.06.004 PubMed PMID: 32561324.

10. Wilkinson-Smith V, Dellschaft N, Ansell J, Hoad C, Marciani L, Gowland P, et al. Mechanisms underlying effects of kiwifruit on intestinal function shown by MRI in healthy volunteers. Aliment Pharmacol Ther. 2019 Mar 1;49(6):759–68. doi:10.1111/apt.15127 PubMed PMID: 30706488.

11. Gunn D, Murthy R, Major G, Wilkinson-Smith V, Hoad C, Marciani L, et al. Contrasting effects of viscous and particulate fibers on colonic fermentation in vitro and in vivo, and their impact on intestinal water studied by MRI in a randomized trial. Am J Clin Nutr. 2020 Sep 1;112(3):595–602. doi:10.1093/AJCN/NQAA173 PubMed PMID: 32619212.

12. Bendezú RA, Mego M, Monclus E, Merino X, Accarino A, Malagelada JR, et al. Colonic content: effect of diet, meals, and defecation. Neurogastroenterology and Motility. 2017 Feb 1;29(2):e12930. doi:10.1111/NMO.12930;WEBSITE:WEBSITE:PERICLES;WGROUP:STRING:PUBLICATION PubMed PMID: 27545449.

13. Aliyu A, Dellschaft N, Hoad C, Williams H, Gaudoin E, Sulaiman S, et al. Magnetic Resonance Imaging Reveals Novel Insights into the Dual Mode of Action of Bisacodyl: A Randomized, Placebo-controlled Trial in Constipation. Clin Pharmacol Ther. 2025 May 1;117(5):1284–91. doi:10.1002/CPT.3532;PAGEGROUP:STRING:PUBLICATION PubMed PMID: 39679695.

14. Bolvig E, Gerdt D, Elise Møller L, Brøndum J, Mohr A. Aalborg Universitet Effects of opium tincture on gastrointestinal function and motility in healthy volunteers: A magnetic resonance imaging study [Internet]. 2024. doi:10.1111/nmo.14941

15. Pritchard SE, Marciani L, Garsed KC, Hoad CL, Thongborisute W, Roberts E, et al. Fasting and postprandial volumes of the undisturbed colon: normal values and changes in diarrhea-predominant irritable bowel syndrome measured using serial MRI. Neurogastroenterology & Motility. 2014 Jan 1;26(1):124–30. doi:10.1111/NMO.12243 PubMed PMID: 24131490.

16. Sandberg TH, Nilsson M, Poulsen JL, Gram M, Frøkjær JB, Østergaard LR, et al. A novel semi-automatic segmentation method for volumetric assessment of the colon based on

magnetic resonance imaging. Abdominal Imaging 2015 40:7. 2015 Jun 9;40(7):2232–41. doi:10.1007/s00261-015-0475-z PubMed PMID: 26054979.

17. Chen X, Williams BM, Vallabhaneni SR, Czanner G, Williams R, Zheng Y. Learning active contour models for medical image segmentation. Proceedings of the IEEE Computer Society Conference on Computer Vision and Pattern Recognition. 2019 Jun 1;2019-June:11624–32. doi:10.1109/CVPR.2019.01190

18. Zhang Z, Keles E, Durak G, Taktak Y, Susladkar O, Gorade V, et al. Large-scale multi-center CT and MRI segmentation of pancreas with deep learning. Med Image Anal. 2025 Jan 1;99:103382. doi:10.1016/J.MEDIA.2024.103382 PubMed PMID: 39541706.

19. Raab F, Strotzer Q, Stroszczynski C, Fellner C, Einspieler I, Haimerl M, et al. Automatic segmentation of liver structures in multi-phase MRI using variants of nnU-Net and Swin UNETR. Scientific Reports 2025 15:1. 2025 Jul 16;15(1):25740-. doi:10.1038/s41598-025-07084-5 PubMed PMID: 40670420.

20. Krishnan C, Schmidt E, Onuoha E, Mrug M, Cardenas CE, Kim H. nnUNet for Automatic Kidney and Cyst Segmentation in Autosomal Dominant Polycystic Kidney Disease. Curr Med Imaging. 2024 Feb 7;20(1):1–9. doi:10.2174/0115734056272767231130110017 PubMed PMID: 38389364.

21. Jaitner N, Ludwig J, Meyer T, Boehm O, Anders M, Huang B, et al. Automated liver and spleen segmentation for MR elastography maps using U-Nets. Sci Rep. 2025 Dec 1;15(1):10762-. doi:10.1038/S41598-025-95157-W;SUBJMETA PubMed PMID: 40155744.

22. Akinci D'antonoli T, Lucas •, Berger K, Indrakanti AK, Vishwanathan N, Weiss J, et al. TotalSegmentator MRI: Robust Sequence-independent Segmentation of Multiple Anatomic Structures in MRI [Internet]. 2025. doi:10.1148/radiol.241613

23. Benson E, Rier L, Millican I, Pritchard S, Costigan C, Pound M, et al. The effect of depth context in the segmentation of the colon in MRI volumes [Internet]. doi:10.1101/2020.03.06.20027722

24. Sharma N, Gupta S, Almogren A, Bharany S, Altameem A, Rehman AU. Advanced gastrointestinal tract organ differentiation using an integrated swin transformer U-Net model for cancer care. Front Phys. 2024 Dec 12;12:1478750. doi:10.3389/fphy.2024.1478750

25. Sharma N, Gupta S, Elkamchouchi DH, Bharany S. Encoder–Decoder Variant Analysis for Semantic Segmentation of Gastrointestinal Tract Using UW-Madison Dataset. Bioengineering 2025, Vol 12,. 2025 Mar 17;12(3). doi:10.3390/bioengineering12030309

26. Nuruzzaman Nobel S, Faruque Sifat O, Islam MR, Sayeed MS, Amiruzzaman M. ResECA-Unet: Enhancement of GI Tract Segmentation Using an Improved U-Net Framework [Internet]. 2024 Mar 29. doi:10.20944/preprints202403.1833.v1

27. Zhang Y, Gong Y, Cui D, Li X, Shen X. DeepGI: An Automated Approach for Gastrointestinal Tract Segmentation in MRI Scans.

28. Zhong Z, Huang L, Feng ST, Lin H, Wang X, Lu B, et al. A comprehensive dataset of magnetic resonance enterography images with intestinal segment annotations. Scientific Data 2025 12:1. 2025 Mar 11;12(1):425-. doi:10.1038/s41597-025-04760-z PubMed PMID: 40069172.

29. Orellana B, Navazo I, Brunet P, Monclús E, Bendezú Á, Azpiroz F. Automatic colon segmentation on T1-FS MR images. Computerized Medical Imaging and Graphics. 2025 Jul 1;123(5):102528. doi:10.1016/j.compmedimag.2025.102528 PubMed PMID: 40112651.

30. Bareja R, Ismail M, Martin D, Nayate A, Yadav I, Labbad M, et al. nnU-Net–based Segmentation of Tumor Subcompartments in Pediatric Medulloblastoma Using Multiparametric MRI: A Multi-institutional Study. https://doi.org/101148/ryai230115. 2024 Aug 21;6(5). doi:10.1148/RYAI.230115

31. Joshi S, Pant M, Malhotra A, Deep K, Snasel V. A nnU-Net-based automatic segmentation of FCD type II lesions in 3D FLAIR MRI images. Front Artif Intell. 2025 Jun 27;8:1601815. doi:10.3389/FRAI.2025.1601815/TEXT

32. Zhou Y, Lalande A, Chevalier C, Baude J, Aubignac L, Boudet J, et al. Deep learning application for abdominal organs segmentation on 0.35 T MR-Linac images. Front Oncol. 2023 Jan 8;13:1285924. doi:10.3389/FONC.2023.1285924/TEXT

33. Ronneberger O, Fischer P, Brox T. 2015-U-Net. ArXiv [Internet]. 2015 [cited 2026 Jun 3];1–8. Available from: http://lmb.informatik.uni-freiburg.de/%0Aarxiv:1505.04597v1

34. Jain I, Willems S, Latre S, De Schepper T. On-the-Fly Data Augmentation for Brain Tumor Segmentation [Internet]. 2025 Sep 29 [cited 2026 Jun 3]. Available from: https://arxiv.org/pdf/2509.24973

35. Li X, Xu ZQJ, Ren Y, Qiu T, Wang X. Towards Reliable Pediatric Brain Tumor Segmentation: Task-Specific nnU-Net Enhancements [Internet]. 2025 Nov 1 [cited 2026 Jun 3]. Available from: https://arxiv.org/pdf/2511.00449

36. Yi Y, Zhuang Q, Xu ZQJ, Wang X, Ren Y, Qiu T. Frequency-Aware Ensemble Learning for BraTS 2025 Pediatric Brain Tumor Segmentation [Internet]. 2025 Sep 18 [cited 2026 Jun 3]. Available from: https://arxiv.org/pdf/2509.19353

37. Neele Dellschaft, Alaa Alhasani, Joshua Reid, Abi Spicer, Caroline Hoad, Luca Marciani, et al. Ingesting methylcellulose fibre gels are as effective as psyllium in reducing colonic fermentation of inulin - a first step to IBS symptom relief. In: International Society for Magnetic Resonance in Medicine 2025. Hawaii: ISMRM; 2025. p. 4927.

38. Eggers H, Brendel B, Duijndam A, Herigault G. Dual-echo Dixon imaging with flexible choice of echo times. Magn Reson Med. 2011;65(1):96–107. doi:10.1002/MRM.22578 PubMed PMID: 20860006.

39. McAuliffe MJ, Lalonde FM, McGarry D, Gandler W, Csaky K, Trus BL. Medical Image Processing, Analysis & Visualization In Clinical Research. IEEE COMPUTER-BASED MEDICAL SYSTEMS (CBMS). 2001;381–6.

40. Yushkevich PA, Piven J, Hazlett HC, Smith RG, Ho S, Gee JC, et al. User-guided 3D active contour segmentation of anatomical structures: Significantly improved efficiency and reliability. Neuroimage. 2006 Jul 1;31(3):1116–28. doi:10.1016/J.NEUROIMAGE.2006.01.015 PubMed PMID: 16545965.

41. Sharma N, Gupta S, Almogren A, Bharany S, Altameem A, Rehman AU. Advanced gastrointestinal tract organ differentiation using an integrated swin transformer U-Net model for cancer care. Front Phys. 2024 Dec 12;12:1478750. doi:10.3389/FPHY.2024.1478750/TEXT

42. Zhang Y, Gong Y, Cui D, Li X, Shen X. DeepGI: An Automated Approach for Gastrointestinal Tract Segmentation in MRI Scans.

43. Plocharski M, Hvaas GB, Møller MG, Thostrup NS, Samuelsen F, Mark EB. Fully Automated Colon Delineation and Volume Estimation in T2-Weighted MRI with a 2D U-Net. Stud Health Technol Inform. 2026 May 21;336:37–41. doi:10.3233/SHTI260104 PubMed PMID: 42174781.

44. Zhang Z, Keles E, Durak G, Taktak Y, Susladkar O, Gorade V, et al. Large-scale multi-center CT and MRI segmentation of pancreas with deep learning. Med Image Anal. 2025 Jan 1;99:103382. doi:10.1016/J.MEDIA.2024.103382 PubMed PMID: 39541706.

45. Aliyu A, Dellschaft N, Hoad C, Williams H, Gaudoin E, Sulaiman S, et al. Magnetic Resonance Imaging Reveals Novel Insights into the Dual Mode of Action of Bisacodyl: A Randomized, Placebo-controlled Trial in Constipation. Clin Pharmacol Ther. 2024 May 1;117(5):1284. doi:10.1002/CPT.3532 PubMed PMID: 39679695.